\documentclass[aps,prl,twocolumn,showpacs,superscriptaddress,groupedaddress]{revtex4}  
\usepackage{graphicx}  
\usepackage{dcolumn}   
\usepackage{bm,amsmath}        
\usepackage{amssymb}   
\usepackage{color}

\newcommand{\beq}{\begin{equation}} 
\newcommand{\eeq}{\end{equation}} 
\newcommand{\bea}{\begin{eqnarray}} 
\newcommand{\eea}{\end{eqnarray}}

\begin{document}


\title{Edge Detecting New Physics the Voronoi Way}
\author{Dipsikha~Debnath} \affiliation{Physics Department,
  University of Florida, Gainesville, FL 32611, USA}
\author{James~S.~Gainer} \affiliation{Physics Department, University
  of Florida, Gainesville, FL 32611, USA}
\author{Doojin~Kim} \affiliation{Physics Department,
  University of Florida, Gainesville, FL 32611, USA}
\author{Konstantin~T.~Matchev} \affiliation{Physics Department,
  University of Florida, Gainesville, FL 32611, USA}
\date{June 12, 2015}
\begin{abstract}
We point out that interesting features in high energy physics data 
can be determined from properties of Voronoi tessellations 
of the relevant phase space.
For illustration, we focus on the detection of kinematic ``edges" in two dimensions,
which may signal physics beyond the standard model. 
After deriving some useful geometric results for Voronoi tessellations 
on perfect grids, we propose several algorithms for tagging the Voronoi cells 
in the vicinity of kinematic edges in real data. We show that the efficiency 
is improved by the addition of a few Voronoi relaxation steps via Lloyd's method.  
By preserving the maximum spatial resolution of the data, Voronoi methods
can be a valuable addition to the data analysis toolkit at the LHC.
\end{abstract}
\maketitle

{\bf Introduction.} 
Experimental searches for new physics (NP) are ultimately searches for
``features'' in the data. In high energy physics, the data is represented by
a collection of ``events", which are distributed in phase space, ${\cal P}$, 
according to the fully differential cross-section
\beq
\frac{d\sigma}{d\vec{\bm x}} \equiv f(\vec{\bm x}, \left\{\alpha\right\}).
\label{fdef}
\eeq
Here $\vec{\bm x}\in {\cal P}$ is a particular phase space point,
typically labelled by momentum components of final state particles,
while $\left\{\alpha\right\}$ is a set of model parameters, e.g., 
particle masses, widths, couplings, etc.  
Thus, the distribution of events 
in phase space is nothing but a Monte Carlo sampling
 of the function (\ref{fdef}),
which generally consists of two contributions:
\beq
f(\vec{\bm x}, \left\{\alpha\right\}) \equiv f_{SM} (\vec{\bm x}, \left\{\alpha_{SM}\right\})
+f_{NP}(\vec{\bm x}, \left\{\alpha_{NP}\right\}).
\label{f}
\eeq 
Here $f_{SM}$ is the distribution expected from Standard Model (SM) 
processes, a.k.a.~``the background",
while $f_{NP}$ describes possible new physics, i.e., ``the signal". 

The traditional method to search for new physics is in
counting experiments, where one measures the total number 
of events in a suitably chosen region of phase space, ${\cal P}_0$.
New physics then manifests itself as an excess over the SM expectation
$\int_{{\cal P}_0} f_{SM} (\vec{\bm x}, \left\{\alpha_{SM}\right\})\, d {\vec{\bm x}}$.
However, a much more powerful approach is to look at the differential 
properties of the observed events in phase space, and attempt to identify 
structural features in their distributions, which might be present in
$f_{NP}$, but not in $f_{SM}$. An example of this 
method is the bump-hunting technique in resonance searches,
where the Breit-Wigner peak in $f_{NP}$ ``stands out" over the 
smooth background described by $f_{SM}$.

The situation gets much more complicated if some of the decay products
(e.g., neutrinos or dark matter particles) are invisible in the detector.
Even then, the signal distribution $f_{NP}$ often exhibits some special 
features\footnote{In general, such kinematic singularities arise in the course of projecting
the full phase space onto a lower-dimensional space of observables
\cite{Kim:2009si}.}, e.g., kinematic endpoints \cite{Hinchliffe:1996iu,Cho:2009ve,Barr:2010zj,Barr:2011xt}, 
kinematic boundaries \cite{Costanzo:2009mq,Burns:2009zi,Matchev:2009iw,Matchev:2009ad,Agrawal:2013uka}, 
kinks \cite{Cho:2007qv,Gripaios:2007is,Barr:2007hy,Cho:2007dh,Burns:2008va} 
and cusps \cite{Han:2009ss,Agashe:2010gt,Han:2012nm,Han:2012nr}, which are {\em absent} 
in the background distribution $f_{SM}$.
This greatly motivates the development of new methods for
reliable identification of such features in the data, which would be
tantamount to a new physics discovery.

In this letter, we focus on two-dimensional high energy particle physics 
data, leaving the straightforward generalization to higher dimensions to a future study
\cite{us}. We assume that the signal distribution exhibits a
characteristic ``edge", along which $f_{NP}$ is discontinuous, 
as is typical of a kinematic boundary in phase space.
Edge detection is a well-studied problem in the
experimental and observational sciences \cite{edgedetection}.  
There are, however, several challenges for edge detection in particle
physics that may frustrate standard approaches, namely

{\em 1. The data may be relatively sparse.}  Most work in edge
  detection has focused on images, where there is a data point at each
  pixel.  By contrast, in particle physics we may want to discover new
  physics with a relatively small number of signal events, when
  large regions of phase space may remain unpopulated.

{\em 2. We may not know analytically the class of distributions, $f_{SM}+f_{NP}$,
   that describe the data.}  If we know the parametric form of the distribution (\ref{f}), 
   likelihood methods can be used to determine edges.
  However, it is generally difficult to obtain an exact analytical form for $f_{SM}$,
  particularly in the case of reducible backgrounds, where detector effects play a major role. 
  We may also wish to be sensitive to ``surprises'' in the data 
  with regards to new physics --- after all, we cannot be sure, 
  {\em \'{a} priori}, that we have correctly guessed the specific 
  new physics model \cite{Debnath:2014eaa}.
  Even if we have some idea of where the new physics edges may be found, 
  a general procedure may still be of greater practical use.

{\em 3. The data may be in more than two dimensions.}  As we
  mentioned above, edge detection is generally applied to
  two-dimensional images.  However, multivariate
  analyses~\cite{multivariate} are ubiquitous in particle physics; 
  in general, we will face the problem of finding an $(n-1)$-dimensional 
  kinematic boundary in an $n$-dimensional parameter space.

The class of methods we propose can handle all three of these
challenges, making them an important addition to the experimentalist's
toolkit for Run 2 at the CERN Large Hadron Collider (LHC).
The starting point of our analysis is the Voronoi
tessellation\footnote{A Voronoi tessellation~\cite{voronoi} is a procedure, 
originally proposed by Dirichlet~\cite{dirichlet}, 
through which a volume containing data points $\{d_i\}$
is divided into regions, $\mathcal{R}_i$, such that each $\mathcal{R}_i$
contains exactly one data point, $d_i$, and for any point $p \in
\mathcal{R}_i$, $d_i$ is the nearest data point.} of our two-dimensional data, where each ``event" 
$i$ is treated as the corresponding generator point for the $i$-th Voronoi polygon \cite{VT}. 
Voronoi tessellations have been successfully used in various 
areas of science, including condensed matter physics~\cite{bowick},
astronomy~\cite{Elyiv:2008bi}
and astrophysics~\cite{SoaresSantos:2010xj,Gerke:2012qq},
but so far have not been applied to data analysis in high energy physics.\footnote{We 
also note the existence of efficient codes for finding
Voronoi tessellations in the form of the {\tt qHULL} algorithms~\cite{qhull}.
Wrappers that allow the use of these algorithms in many frameworks also exist.  
}
At the same time, one can think of the outcome from any high-energy 
collider experiment as an inhomogeneous Poisson point process 
with an intensity function (\ref{fdef}). Voronoi methods were designed for the 
analysis of precisely such processes.

\begin{figure}[t] 
\includegraphics[height=3.6cm]{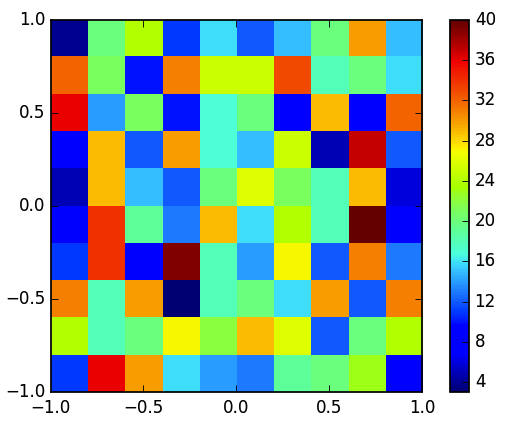} 
\includegraphics[height=3.6cm]{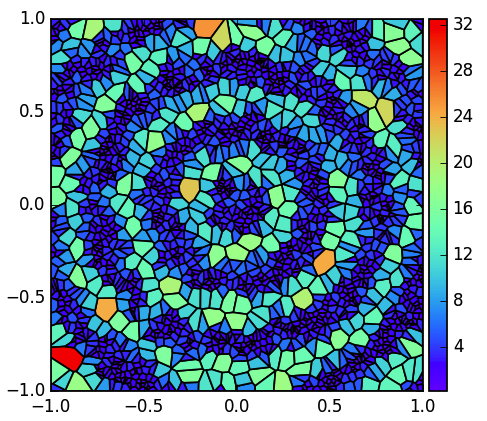}
\caption{ Left: a traditional two-dimensional binned histogram
for a planar stochastic point process with 2000 events generated 
according to the distribution (\ref{fsin}). Right: the Voronoi tessellation of the same data.
\label{fig:binvor} }
\end{figure}

The Voronoi approach is ideally suited for finding interesting (e.g., singular) features
in $f_{NP}$, since it preserves the maximum
spatial resolution in the data~\cite{Cappellari:2009sc}. To see this,
consider a toy example where 2000 data ``points" $(x,y)$
are generated according to the periodic function
\beq
f(\vec{\bm x}) = 1 + \sin \left(6\pi \sqrt{x^2+y^2} \right)\, .
\label{fsin}
\eeq
The standard approach is to bin the data, e.g., as shown in the left panel of Fig.~\ref{fig:binvor}.
It is clear that interesting features of the underlying distribution are being lost as a result of 
averaging within each bin. In contrast, the Voronoi tessellation of the same data,
shown in the right panel of Fig.~\ref{fig:binvor}, clearly displays the radial periodicity 
and rotational symmetry of the data. Furthermore, by construction, the areas $a_i$ 
of the Voronoi polygons serve as local estimators of the values of the generating function $f(\vec{\bm x})$
at the location $\vec{\bm x}_i$ of each generator point $p_i$
\beq
f(\vec{\bm x}_i) \sim 1/a_i. 
\label{eq:areascaling}
\eeq
One can further interpolate between the generator points by the method
of natural neighbor interpolation~\cite{VT}.

{\bf Voronoi Methods for Edge Detection.} Since we do not assume the exact knowledge
of $f_{NP}$, we do not attempt here to reconstruct the function itself, but focus instead on
finding edge features such as discontinuities \cite{1d}. 
Edge detection algorithms for binned data readily exist \cite{canny};
our methods will apply to the corresponding Voronoi tessellation
and include the following steps:
\begin{enumerate}
\item Construct the Voronoi tessellation for the data set.
\item Compute relevant attributes of the Voronoi cells.
\item (Optionally) use the information from the previous step 
to further process the data in some way.
\item Use some criterion to flag ``candidate'' edge cells.
\end{enumerate}

\begin{figure}[t!] 
\includegraphics[width=0.49\columnwidth]{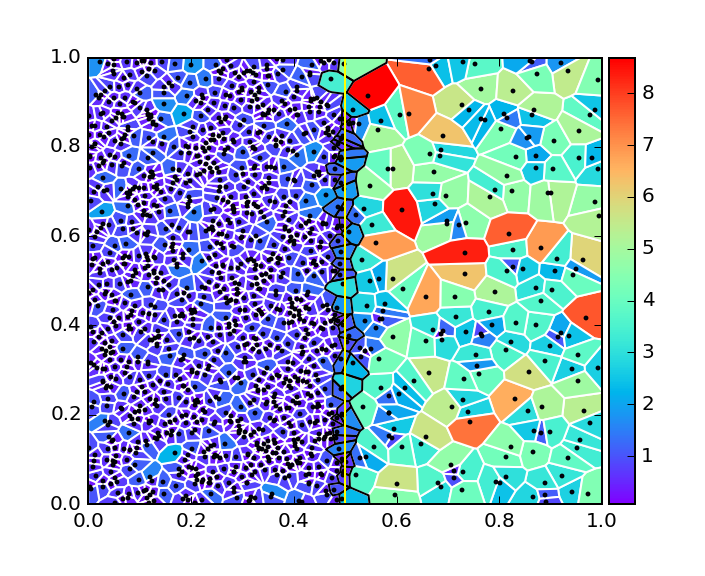} 
\includegraphics[width=0.49\columnwidth]{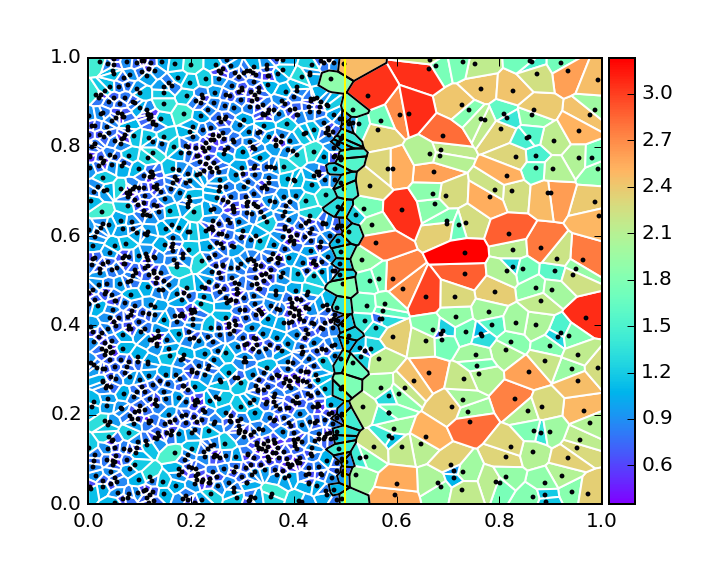}\\
\includegraphics[width=0.49\columnwidth]{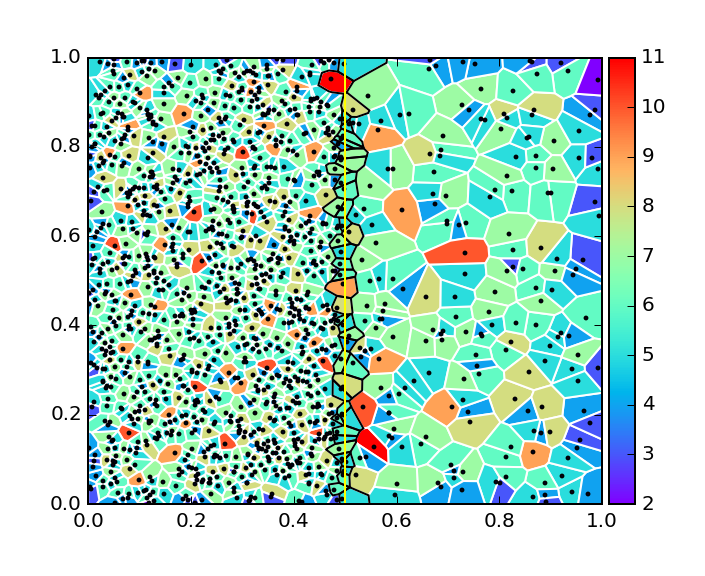} 
\includegraphics[width=0.49\columnwidth]{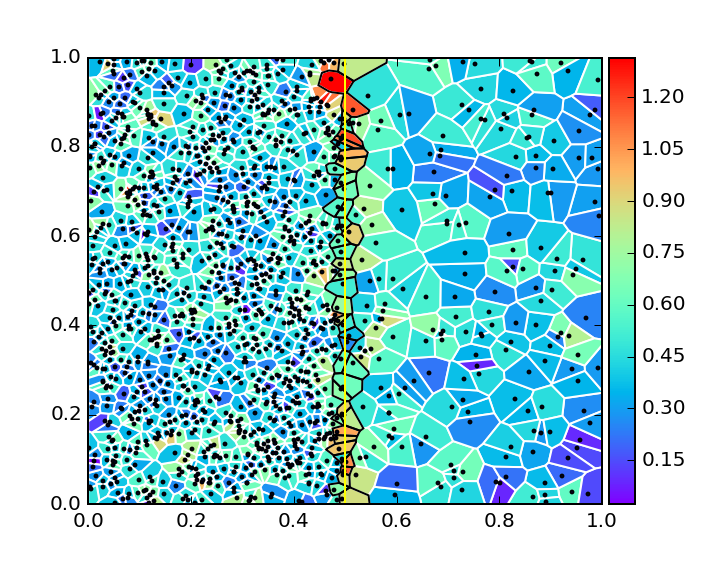}
\caption{ The Voronoi tessellation of 1400 data points distributed according to the 
probability density (\ref{fstep}) with $\rho=6$. The Voronoi polygons have been color-coded by
their area (upper left), perimeter (upper right), number of neighboring polygons (lower left)
or scaled variance (\ref{defvar}) (lower right). 
\label{fig:temp} }
\end{figure}

We can gain useful intuition from a toy example illustrating this procedure.
Consider the probability distribution within the unit square,
\beq
f(x,y) = 
\frac{2}{1+\rho}
\left[ \rho H(0.5-x) + H(x-0.5) \right] ,
\label{fstep}
\eeq
where $H(x)$ is the Heaviside step function and $\rho$ is a constant density ratio. 
We generate 1400 points and show the resulting Voronoi tessellation in Fig.~\ref{fig:temp},
where the Voronoi polygons have been color-coded according to some standard properties,
e.g., area, perimeter, or number of immediate neighbors.
Our goal will be to investigate the vertical edge at $x=0.5$ (yellow solid line), 
which divides the square into left (L) and right (R) regions of constant, but unequal densities.\footnote{For 
the convenience of the reader, in Fig.~\ref{fig:temp} 
the Voronoi cells crossed by the edge at $x=0.5$ are outlined in black,
while the boundaries of the remaining Voronoi cells away from the edge are kept white.}

The first three panels of Fig.~\ref{fig:temp} reveal that: 
(a) the areas of the Voronoi polygons scale in accordance with (\ref{eq:areascaling});
(b) the perimeter is somewhat correlated with the area, but the connection is not as direct as in (\ref{eq:areascaling});
(c) the typical number of neighbors is similar in the two bulk regions.
Neither of these three quantities is able to characteristically pick out the edge cells
(outlined in black). This motivates us to introduce the scaled variance of the areas of the neighboring cells, 
\beq
\frac{\sigma_a}{\bar{a}} \equiv \frac{1}{\bar{a}}\, \sqrt{\sum_{n\in N_i} \frac{\left(a_n-\bar{a}\right)^2}{|N_i|-1}},
\label{defvar}
\eeq
where $N_i$ is the set of neighbors of the $i$-th Voronoi polygon,
and $\bar{a}(N_i)$ is their mean area. The lower right panel in 
Fig.~\ref{fig:temp} shows that the scaled variance is quite successful 
in identifying edge cells, thus we shall choose (\ref{defvar}) as our 
main selection variable\footnote{This is not the only option, however --- we have
investigated a number of other promising variables which will be discussed
in a longer publication \cite{us}.}.

\begin{figure}[t] 
\includegraphics[height=3.2cm]{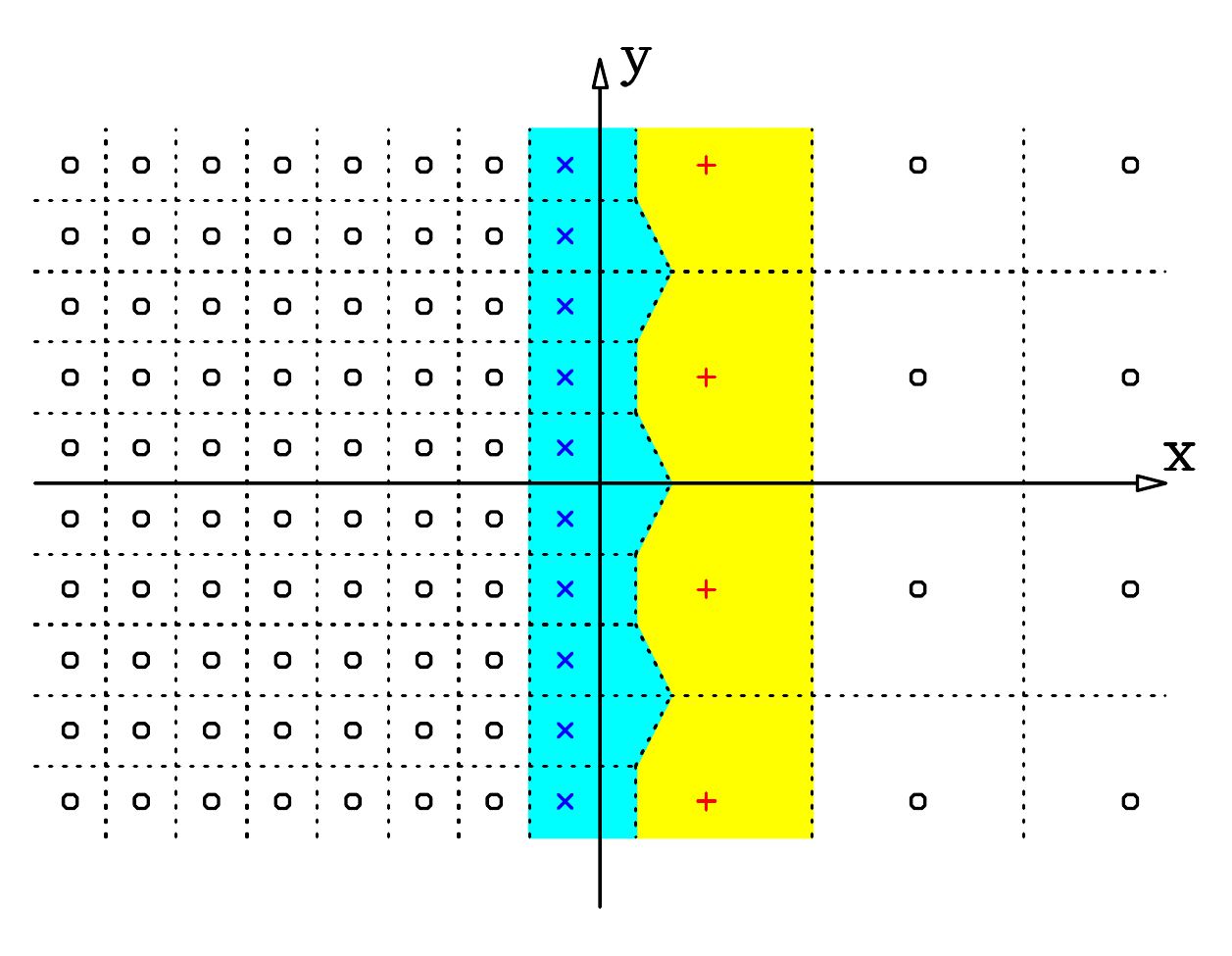}
\includegraphics[height=3.2cm]{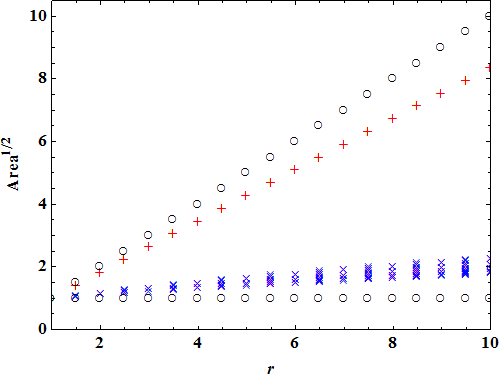} \\
\includegraphics[height=3.23cm]{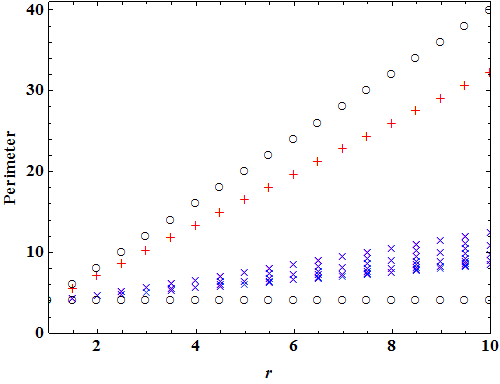}
\includegraphics[height=3.2cm]{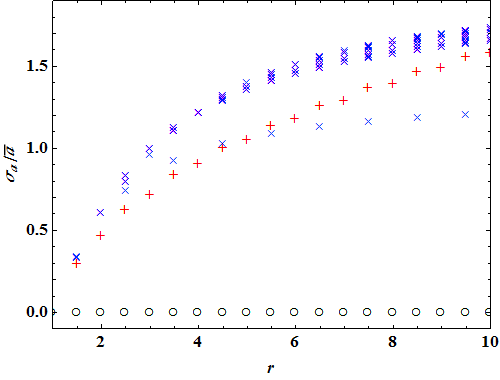} 
\caption{ A perfect lattice grid (\ref{lattice}) for $r=3$ (upper left), and dependence on $r$ for
several parameters of interest: cell area (upper right), cell perimeter (lower left) and scaled variance (lower right). 
Black circles denote bulk cells, while blue $\times$ (red $+$) symbols
indicate edge cells in the L (R) region.
\label{fig:reg} }
\end{figure}

The results from the stochastic process shown in Fig.~\ref{fig:temp},
and in particular the success of the variable in (\ref{defvar}),
can be understood analytically if we analyze a perfect lattice 
which mimics the probability distribution (\ref{fstep}). 
The grid is generated by two integers, $n$ and $m$, as
\beq
\vec{R} = \left[ \left( n+0.5 \right) \hat x + \left( m+0.5\right) \hat y\right]
\left[H(-n) + rH(n)\right],
\label{lattice}
\eeq
where the vectors $\hat x$ and $\hat y$ form an orthonormal basis and $r\equiv\sqrt{\rho}$
is the corresponding linear density ratio.
An example grid for $r=3$ is shown in the upper left panel of Fig.~\ref{fig:reg}.
In particular, we focus on the two columns of edge cells:
in the L region (blue $\times$ symbols) and the R region (red $+$ symbols).
The remaining three panels in Fig.~\ref{fig:reg} show plots of
some of their properties as a function of $r$.
We see that both the area and the perimeter of the edge cells are 
slightly modified from their nominal values in the bulk, but remain 
in between the two extreme bulk values. On the other hand, 
the scaled variance is identically zero for both bulk regions,
and monotonically increasing with $r$ for the edge region. 

\begin{figure}[t] 
\includegraphics[width=0.49\columnwidth]{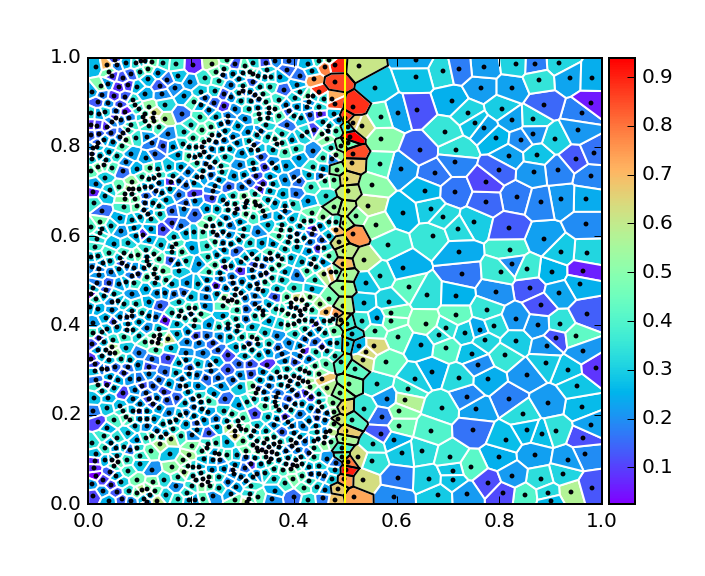} 
\includegraphics[width=0.49\columnwidth]{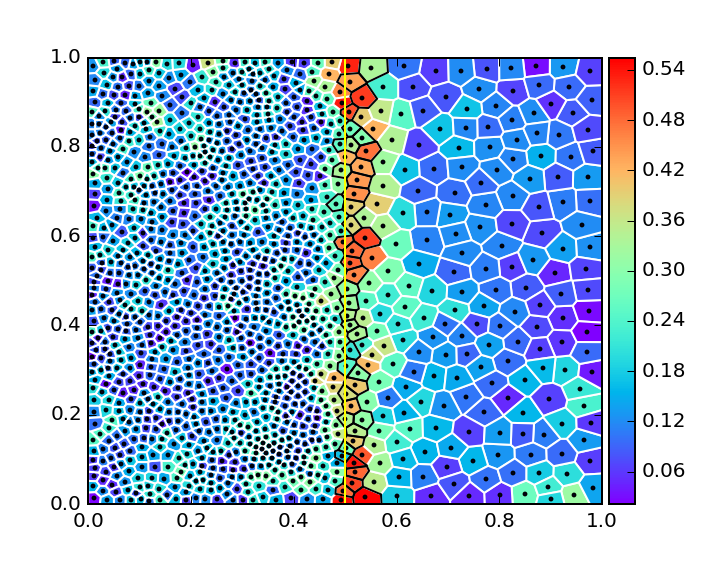}
\caption{ Evolution of the Voronoi tessellation from Fig.~\ref{fig:temp}
after one (left panel) and five (right panel) Lloyd relaxation steps.
The cells are color coded by the scaled variance (\ref{defvar}). 
\label{fig:movie} }
\end{figure}

{\bf Voronoi relaxation via Lloyd's algorithm.} 
Since we are dealing with a stochastic process, statistical fluctuations are inevitably present in the data.
In particular, the lower right panel in Fig.~\ref{fig:temp} reveals pockets of bulk cells with relatively high 
values of $\sigma_a/\bar{a}$. Here we propose to filter out such extraneous cells by first applying a few
iterations of Lloyd's algorithm \cite{lloyd}, where at each iteration, the generator point is moved to the centroid of
the corresponding Voronoi cell.\footnote{An alternative approach, illustrated below in Fig.~\ref{fig:susy},
would be to leave the original Voronoi tessellation 
intact, but extend the calculation of (\ref{defvar}) to include next-to-nearest neighbors, next-to-next-to-nearest neighbors, etc.}
The left (right) panel in Fig.~\ref{fig:movie} shows the result of this operation 
after one (five) such Lloyd iterations. As expected, relaxation causes the Voronoi polygons to become more 
regularly shaped\footnote{For a large number of iterations, Lloyd's algorithm converges to a regular hexagonal
grid in the bulk.}. More importantly, the fluctuations within the bulk regions are washed out, 
thus increasing the contrast between edge cells and bulk cells.

Fig.~\ref{fig:movie} reveals that Voronoi relaxation causes a net flow of the data points 
from the dense L region towards the sparse R region. As a result, the edge cells with high $\sigma_a/\bar{a}$
have moved away from their original locations (near the vertical yellow line). This is why, if we were to tag 
edge candidates after a certain number of Lloyd iterations, we need to trace them back 
to their original locations before doing any further quantitative data analysis.

\begin{figure}[b] 
\includegraphics[height=3.1cm]{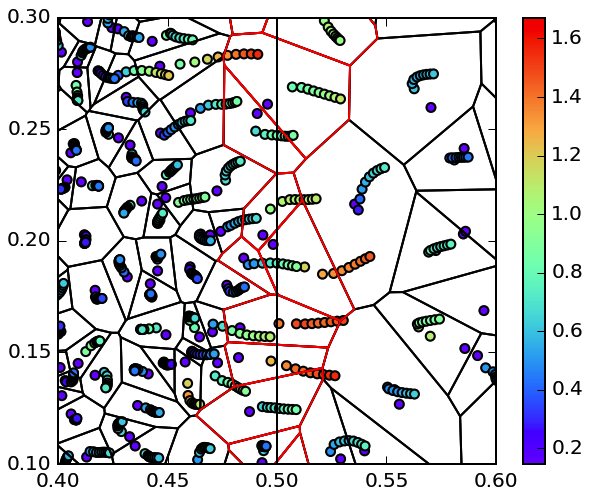} 
\includegraphics[height=3.0cm]{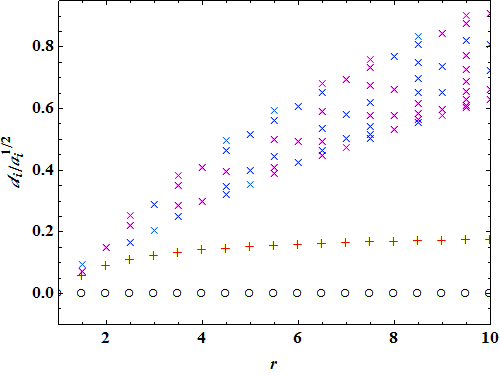}
\caption{ Left: a zoomed-in region near the vertical edge, showing the original 
generator points, and their subsequent locations in the course of several Lloyd iterations.
The points are color-coded according to the scaled displacement, $d_i/\sqrt{a_i}$.
Right: Predictions for $d_i/\sqrt{a_i}$ for the first Lloyd step, as a function of $r$, for
the case of a perfect lattice grid (\ref{lattice}).
\label{fig:worms} }
\end{figure}

The left panel in Fig.~\ref{fig:worms} takes a closer look at one representative area near the edge and 
shows the result of several successive Lloyd iterations. In general, each generator point $i$ is displaced a certain
distance $d_i$ from its original location. It is interesting to note that the edge points appear to be displaced 
the farthest, which is easy to understand in terms of simple diffusion. This observation suggests another criterion
for selecting edge cells --- based on their displacement during the Lloyd relaxation. For convenience, we define
a dimensionless quantity, the scaled displacement $d_i/\sqrt{a_i}$, where we normalize by the square root 
of the cell area, $a_i$. The color coding in the left panel in Fig.~\ref{fig:worms} demonstrates that the scaled displacement
could be a useful alternative to the scaled variance (\ref{defvar}). This is confirmed for the case of the perfect grid (\ref{lattice})
by the exact results shown in the right panel of Fig.~\ref{fig:worms}, . 

\begin{figure}[t] 
\includegraphics[width=0.49\columnwidth]{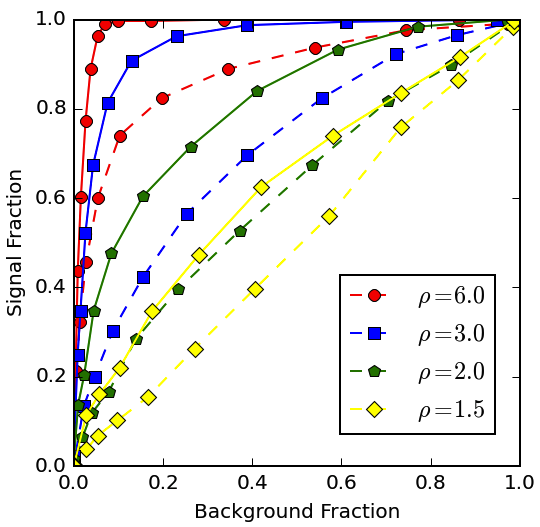} 
\includegraphics[width=0.49\columnwidth]{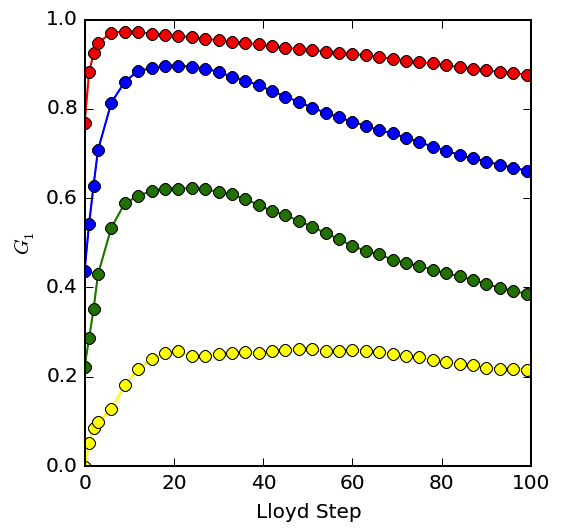}
\caption{ Left: ROC curves $\varepsilon_S(\varepsilon_B)$ using (\ref{defvar}) as the discriminating variable.
Right: The corresponding Gini index (\ref{Gini}) as a function of the number of Lloyd iterations.
\label{fig:ROC} }
\end{figure}

Having defined a selection algorithm for edge cells, it is prudent to study quantitatively the efficiency 
of the algorithm, e.g., in terms of ROC curves \cite{ROC}.   
For this purpose, we generate high statistics samples for (\ref{fstep}), where we treat the 
edge cells as ``signal" and the bulk cells as ``background". We then plot the signal selection efficiency,
$\varepsilon_S$, versus the background efficiency, $\varepsilon_B$, 
for different values of the minimum cut on the variable (\ref{defvar}).
In the left panel of Fig.~\ref{fig:ROC} we show several $\varepsilon_S(\varepsilon_B)$ curves,
for different values of the density ratio $\rho$, and either with (solid) or without (dashed) Lloyd relaxation.
We see that the algorithm works better for higher density contrasts between the two regions.
Note also the significant improvement as a result of adding the Voronoi relaxation.

In order to quantify the accuracy of our selection of edge cells, we use the standard area 
under the curve \cite{AUROC} (AUROC) as represented by the Gini coefficient
\beq
G_1\equiv 2\, {\rm AUROC} - 1 = 2\int_0^1 d\varepsilon_B \times \varepsilon_S(\varepsilon_B) -1,
\label{Gini}
\eeq
where a value of $1$ is obtained from the ROC curve of a perfectly discriminating variable, while
a value of $0$ corresponds to a totally random selection of events. The right panel of Fig.~\ref{fig:ROC}
shows the dependence of $G_1$ on the number of Lloyd steps. We see that the accuracy 
improves very quickly within the first few iterations, and reaches an optimum plateau, after which the
power of the test is degraded as the Voronoi grid begins to asymptote to the centroidal tessellation.

\begin{figure}[t] 
\includegraphics[height=3.3cm]{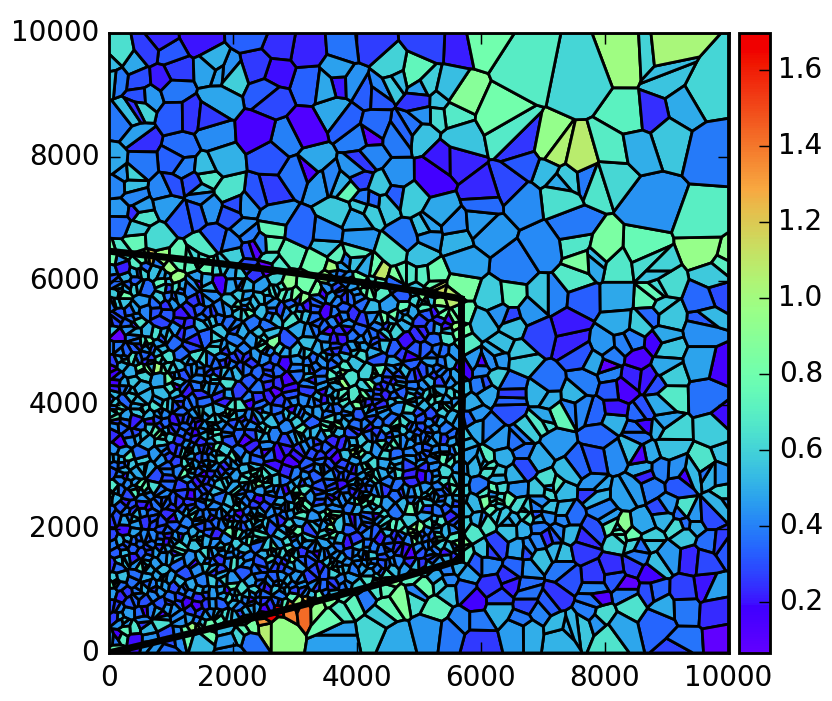}
\includegraphics[height=3.3cm]{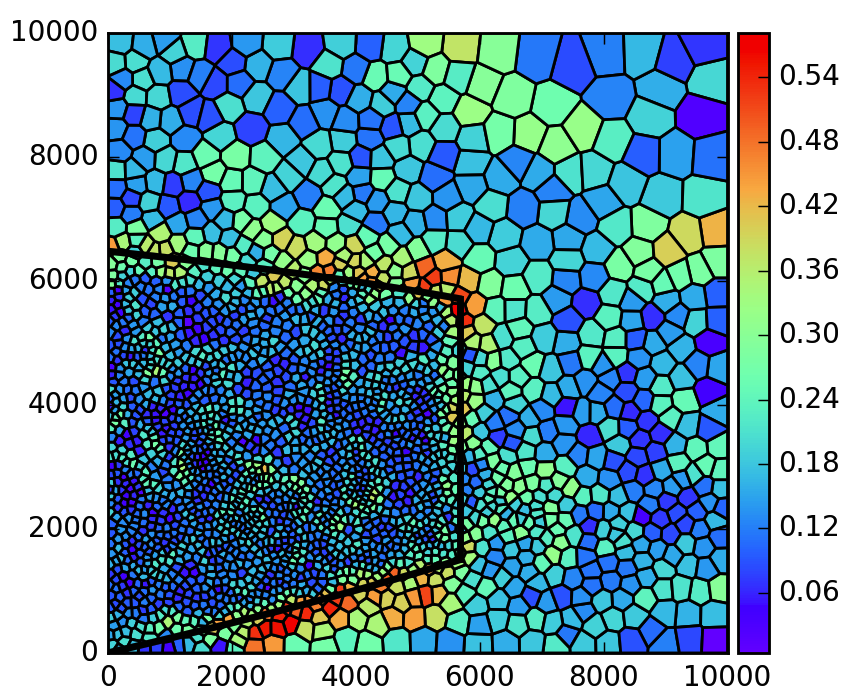}\\ 
\includegraphics[height=3.3cm]{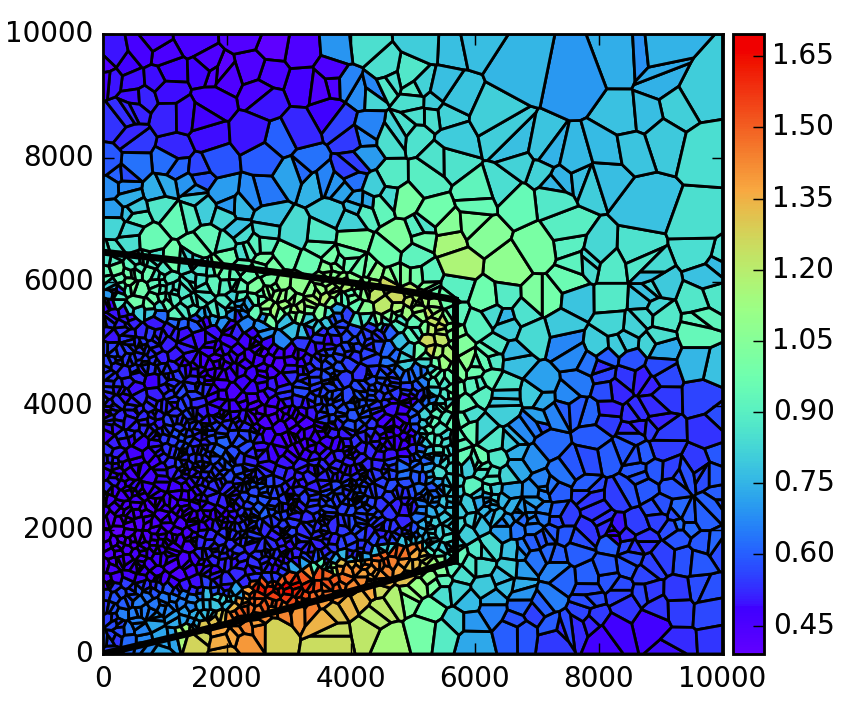}
\includegraphics[height=3.3cm]{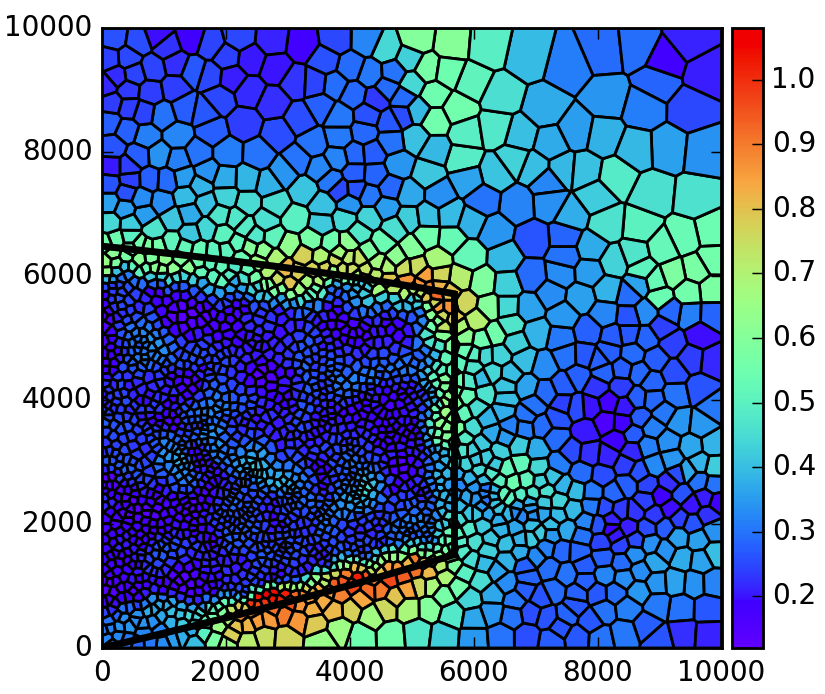}
\caption{ Various Voronoi tessellations of the data from the supersymmetry example considered in the text.
\label{fig:susy} }
\end{figure}

{\bf An example from supersymmetry.} As an application of the 
proposed edge detection method, we consider a standard benchmark example from supersymmetry,
squark pair production at the 13 TeV LHC. For simplicity, we focus on 
asymmetric events in which one squark undergoes a long cascade decay through 
a heavy neutralino, $\tilde \chi^0_2$; a slepton, $\tilde \ell$; and a light neutralino, $\tilde\chi^0_1$;
while the other decays directly to $\tilde\chi^0_1$. The mass spectrum was chosen as
$m_{\tilde q}=400$ GeV, $m_{\tilde\chi^0_2}=300$ GeV, 
$m_{\tilde \ell}=280$ GeV, and $m_{\tilde\chi^0_1}=200$ GeV.
The observed final state particles are two jets and two leptons,
whose invariant mass distributions have been well studied and 
are known to exhibit kinematic edges. 
Here we focus on the dilepton invariant mass, $m_{\ell\ell}$,
and the three-body jet-lepton-lepton invariant mass, $m_{j\ell\ell}$.
With the correct jet assignment, signal events are 
constrained to the region outlined by the solid black line in Fig.~\ref{fig:susy}
\cite{Lester:2006cf,Matchev:2009iw}, where for plotting convenience we use
the $(m^2_{\ell\ell}, (m^2_{j\ell\ell}-m^2_{\ell\ell})/6)$ plane.
Since we cannot measure the charge of the jet, there is a two-fold 
combinatorial ambiguity, thus the plot contains two entries per event.
We also include the main SM background from $t\bar{t}$ dilepton events.

In the upper two panels of Fig.~\ref{fig:susy}, the Voronoi cells 
are color coded by their scaled variance (\ref{defvar}).
The left panel is the original data, while the right panel
includes 5 Lloyd steps. In the lower left panel we reconsider the original data, but
extend the calculation of (\ref{defvar}) to include up to 5 tiers of nearest neighbors.
We see that both Voronoi relaxation as well as including more tiers of neighbors 
have the benefit of reducing the fluctuations and sharpening the edge. 
The two procedures can also be done simultaneously --- the lower right panel of Fig.~\ref{fig:susy}
shows the result after 3 Lloyd steps and including 3 tiers of neighbors.


{\bf Summary.} In this letter, we have argued that the discovery of new kinematic features is 
an essential step in the discovery of physics beyond the standard model and
have advocated the use of Voronoi methods for this purpose.
The great flexibility of Voronoi methods is a blessing for the experimentalist; 
many useful properties of the Voronoi cells can be used to construct powerful variables
tailored to specific new physics scenarios. A voluminous, quantitative study of the
many options available to the experimenter will be presented in a companion paper~\cite{us}.

{\bf Acknowledgements.} We thank S.~Das, C.~Kilic, Z.~Liu, R.~Lu, P.~Ramond, X.~Tata, 
J.~Thaler, B.~Tweedie, and D.~Yaylali for useful discussions.
Work supported in part by U.S. Department of Energy Grant
DE-SC0010296.  DK acknowledges support by LHC-TI postdoctoral fellowship 
under grant NSF-PHY-0969510.


\begin{thebibliography}{99}

\bibitem{Kim:2009si} 
  I.~W.~Kim,
  ``Algebraic Singularity Method for Mass Measurement with Missing Energy,''
  Phys.\ Rev.\ Lett.\  {\bf 104}, 081601 (2010)
  [arXiv:0910.1149 [hep-ph]].

\bibitem{Hinchliffe:1996iu} 
  I.~Hinchliffe, F.~E.~Paige, M.~D.~Shapiro, J.~Soderqvist and W.~Yao,
  ``Precision SUSY measurements at CERN LHC,''
  Phys.\ Rev.\ D {\bf 55}, 5520 (1997)
  [hep-ph/9610544].

\bibitem{Cho:2009ve} 
  W.~S.~Cho, J.~E.~Kim and J.~H.~Kim,
  ``Amplification of endpoint structure for new particle mass measurement at the LHC,''
  Phys.\ Rev.\ D {\bf 81}, 095010 (2010)
  [arXiv:0912.2354 [hep-ph]].
  
\bibitem{Barr:2010zj} 
  A.~J.~Barr and C.~G.~Lester,
  ``A Review of the Mass Measurement Techniques proposed for the Large Hadron Collider,''
  J.\ Phys.\ G {\bf 37}, 123001 (2010)
  [arXiv:1004.2732 [hep-ph]].
  
\bibitem{Barr:2011xt} 
  A.~J.~Barr, T.~J.~Khoo, P.~Konar, K.~Kong, C.~G.~Lester, K.~T.~Matchev and M.~Park,
  ``Guide to transverse projections and mass-constraining variables,''
  Phys.\ Rev.\ D {\bf 84}, 095031 (2011)
  [arXiv:1105.2977 [hep-ph]].

\bibitem{Costanzo:2009mq} 
  D.~Costanzo and D.~R.~Tovey,
  ``Supersymmetric particle mass measurement with invariant mass correlations,''
  JHEP {\bf 0904}, 084 (2009)
  [arXiv:0902.2331 [hep-ph]].
  
\bibitem{Burns:2009zi} 
  M.~Burns, K.~T.~Matchev and M.~Park,
  ``Using kinematic boundary lines for particle mass measurements and disambiguation in SUSY-like events with missing energy,''
  JHEP {\bf 0905}, 094 (2009)
  [arXiv:0903.4371 [hep-ph]].
      
\bibitem{Matchev:2009iw} 
  K.~T.~Matchev, F.~Moortgat, L.~Pape and M.~Park,
  ``Precise reconstruction of sparticle masses without ambiguities,''
  JHEP {\bf 0908}, 104 (2009)
  [arXiv:0906.2417 [hep-ph]].
  
\bibitem{Matchev:2009ad} 
  K.~T.~Matchev and M.~Park,
  ``A General method for determining the masses of semi-invisibly decaying particles at hadron colliders,''
  Phys.\ Rev.\ Lett.\  {\bf 107}, 061801 (2011)
  [arXiv:0910.1584 [hep-ph]].

\bibitem{Agrawal:2013uka} 
  P.~Agrawal, C.~Kilic, C.~White and J.~H.~Yu,
  ``Improved mass measurement using the boundary of many-body phase space,''
  Phys.\ Rev.\ D {\bf 89}, no. 1, 015021 (2014)
  [arXiv:1308.6560 [hep-ph]].
  
\bibitem{Cho:2007qv} 
  W.~S.~Cho, K.~Choi, Y.~G.~Kim and C.~B.~Park,
  ``Gluino Stransverse Mass,''
  Phys.\ Rev.\ Lett.\  {\bf 100}, 171801 (2008)
  [arXiv:0709.0288 [hep-ph]].

\bibitem{Gripaios:2007is} 
  B.~Gripaios,
  ``Transverse observables and mass determination at hadron colliders,''
  JHEP {\bf 0802}, 053 (2008)
  [arXiv:0709.2740 [hep-ph]].

\bibitem{Barr:2007hy} 
  A.~J.~Barr, B.~Gripaios and C.~G.~Lester,
  ``Weighing Wimps with Kinks at Colliders: Invisible Particle Mass Measurements from Endpoints,''
  JHEP {\bf 0802}, 014 (2008)
  [arXiv:0711.4008 [hep-ph]].

\bibitem{Cho:2007dh} 
  W.~S.~Cho, K.~Choi, Y.~G.~Kim and C.~B.~Park,
  ``Measuring superparticle masses at hadron collider using the transverse mass kink,''
  JHEP {\bf 0802}, 035 (2008)
  [arXiv:0711.4526 [hep-ph]].
      
\bibitem{Burns:2008va} 
  M.~Burns, K.~Kong, K.~T.~Matchev and M.~Park,
  ``Using Subsystem MT2 for Complete Mass Determinations in Decay Chains with Missing Energy at Hadron Colliders,''
  JHEP {\bf 0903}, 143 (2009)
  [arXiv:0810.5576 [hep-ph]].
    
\bibitem{Han:2009ss} 
  T.~Han, I.~W.~Kim and J.~Song,
  ``Kinematic Cusps: Determining the Missing Particle Mass at Colliders,''
  Phys.\ Lett.\ B {\bf 693}, 575 (2010)
  [arXiv:0906.5009 [hep-ph]].

\bibitem{Agashe:2010gt} 
  K.~Agashe, D.~Kim, M.~Toharia and D.~G.~E.~Walker,
  ``Distinguishing Dark Matter Stabilization Symmetries Using Multiple Kinematic Edges and Cusps,''
  Phys.\ Rev.\ D {\bf 82}, 015007 (2010)
  [arXiv:1003.0899 [hep-ph]].
  
\bibitem{Han:2012nm} 
  T.~Han, I.~W.~Kim and J.~Song,
  ``Kinematic Cusps With Two Missing Particles I: Antler Decay Topology,''
  Phys.\ Rev.\ D {\bf 87}, no. 3, 035003 (2013)
  [arXiv:1206.5633 [hep-ph]].

\bibitem{Han:2012nr} 
  T.~Han, I.~W.~Kim and J.~Song,
  ``Kinematic Cusps with Two Missing Particles II: Cascade Decay Topology,''
  Phys.\ Rev.\ D {\bf 87}, no. 3, 035004 (2013)
  [arXiv:1206.5641 [hep-ph]].
    
\bibitem{us} 
  D.~Debnath, J.~S.~Gainer, D.~Kim and K.~T.~Matchev,
  ``Voronoi Tesselation Methods in Particle Physics,''
  (work in progress).

\bibitem{edgedetection}
See, e.g., Davies, E.~R., ``Computer \& Machine Vision: Theory, Algorithms, Practicalities",
Academic Press; 4 edition (March 19, 2012).

\bibitem{Debnath:2014eaa} 
  D.~Debnath, J.~S.~Gainer and K.~T.~Matchev,
  ``Discoveries far from the Lamppost with Matrix Elements and Ranking,''
  Phys.\ Lett.\ B {\bf 743}, 1 (2015)
  [arXiv:1405.5879 [hep-ph]].
  
\bibitem{multivariate}
  P.~C.~Bhat,
  ``Multivariate Analysis Methods in Particle Physics,''
  Ann.\ Rev.\ Nucl.\ Part.\ Sci.\  {\bf 61}, 281 (2011).

\bibitem{voronoi}
G.~Voronoi,
``Nouvelles applications des param\`etres continus \`a la th\'eorie des formes quadratiques,"
Journal f\"ur die Reine und Angewandte Mathematik, {\bf 133}, 97 (1908).

\bibitem{dirichlet}
G.~L.~Dirichlet,
"\"Uber die Reduktion der positiven quadratischen Formen mit drei unbestimmten ganzen Zahlen," 
Journal f\"ur die Reine und Angewandte Mathematik, {\bf 40}, 209 (1850).

\bibitem{VT}
See, e.g., S.~Okabe, B.~Boots and K.~Sugihara,
``Spatial Tessellations: Concepts and Applications of Voronoi Diagrams,"
John Wiley \& Sons, 1992.

\bibitem{bowick}
L.~Giomi and M.~Bowick,
``Crystalline Order On Riemannian Manifolds With Variable Gaussian Curvature And Boundary,"
 Phys.\ Rev.\ B {\bf 76}, 054106 (2007)
 [arXiv:cond-mat/0702471].

\bibitem{Elyiv:2008bi} 
  A.~Elyiv, O.~Melnyk and I.~Vavilova,
  ``High-order 3D Voronoi tessellation for identifying Isolated galaxies, Pairs and Triplets,''
  Mon.\ Not.\ Roy.\ Astron.\ Soc.\  {\bf 394}, 1409 (2009)
  [arXiv:0810.5100 [astro-ph]].
    
\bibitem{SoaresSantos:2010xj} 
  M.~Soares-Santos, R.~R.~de Carvalho, J.~Annis, R.~R.~Gal, F.~La Barbera, P.~A.~A.~Lopes, R.~H.~Wechsler and M.~T.~Busha {\it et al.},
  ``The Voronoi Tessellation cluster finder in 2+1 dimensions,''
  Astrophys.\ J.\  {\bf 727}, 45 (2011)
  [arXiv:1011.3458 [astro-ph.CO]].

\bibitem{Gerke:2012qq} 
  B.~F.~Gerke, J.~A.~Newman, M.~Davis, A.~L.~Coil, M.~C.~Cooper, A.~A.~Dutton, S.~M.~Faber and P.~Guhathakurta {\it et al.},
  ``The DEEP2 Galaxy Redshift Survey: The Voronoi-Delaunay Method catalog of galaxy groups,''
  Astrophys.\ J.\  {\bf 751}, 50 (2012)
  [arXiv:1203.3899 [astro-ph.CO]].
  
\bibitem{qhull} 
C.~B.~Barber, D.~P~Dobkin and H.~T.~Huhdanpaa, 
``The Quickhull algorithm for convex hulls,'' 
ACM Trans. on Mathematical Software, 22(4):469-483, 
Dec 1996, http://www.qhull.org.

\bibitem{Cappellari:2009sc} 
  M.~Cappellari,
  ``Voronoi binning: Optimal adaptive tessellations of multi-dimensional data,''
  arXiv:0912.1303 [astro-ph.IM].


\bibitem{1d} 
 For a study in one dimension, see D.~Curtin,
  ``Mixing It Up With MT2: Unbiased Mass Measurements at Hadron Colliders,''
  Phys.\ Rev.\ D {\bf 85}, 075004 (2012)
  [arXiv:1112.1095 [hep-ph]].

\bibitem{canny}
See, e.g., J.~Canny, 
``A Computational Approach To Edge Detection," 
IEEE Trans.~Pattern Analysis and Machine Intelligence, 
8(6):679Ð698, (1986).

\bibitem{lloyd}
S.~P.~Lloyd,
``Least squares quantization in PCM," IEEE Trans.~on Information Theory, 28 (2): 129Ð137,
(1982).

\bibitem{ROC}
T.~Fawcett, ``An introduction to ROC analysis,"
Pattern Recognition Letters, {\bf 27}, Issue 8, 861Ð874, (2006).

 \bibitem{AUROC}
 J.~Hanley and B.~McNeil, 
 ``The Meaning and Use of the Area under a Receiver Operating Characteristic (ROC) Curve,"
Radiology, {\bf 143} (1), 29-36 (1982).

\bibitem{Lester:2006cf} 
  C.~G.~Lester, M.~A.~Parker and M.~J.~White,
  ``Three body kinematic endpoints in SUSY models with non-universal Higgs masses,''
  JHEP {\bf 0710}, 051 (2007)
  [hep-ph/0609298].
   
\end{thebibliography}
\end{document}